\documentclass[usenatbib]{mnras}
\usepackage{graphicx}

\title[iPTF14hls]{Strong late-time circumstellar interaction in the  peculiar supernova iPTF14hls\footnote{Observations reported here were obtained at the MMT Observatory, a joint facility of the University of Arizona and the Smithsonian Institution.} }

\author[Andrews $\&$ Smith 2018]{Jennifer E. Andrews$^1$\thanks{Email:
    jandrews@as.arizona.edu}, {Nathan Smith}$^1$\\
     $^1$Steward Observatory, University of Arizona, 933 North Cherry Avenue, Tucson, AZ 85721, USA \\ }

\date{Accepted XXX. Received YYY; in original form ZZZ}

\pubyear{2018}

\begin{document}
\label{firstpage}
\pagerange{\pageref{firstpage}--\pageref{lastpage}}
\maketitle

\begin{abstract}

  We present a moderate-resolution spectrum of the peculiar Type~II
  supernova iPTF14hls taken on day 1153 after discovery.  This
  spectrum reveals the clear signature of shock interaction with dense
  circumstellar material (CSM). We suggest that this CSM interaction
  may be an important clue for understanding the extremely unusual
  photometric and spectroscopic evolution seen over the first 600 days
  of iPTF14hls.  The late-time spectrum shows a double-peaked
  intermediate-width H$\alpha$ line indicative of expansion speeds
  around 1000 km s$^{-1}$, with the double-peaked shape hinting at a
  disc-like geometry in the CSM.  If the CSM was highly asymmetric,
  perhaps in a disc or torus that was ejected from the star 3-6 years
  prior to explosion, then the CSM interaction could have been overrun
  and hidden below the SN ejecta photosphere from a wide range of
  viewing angles.  In that case, CSM interaction luminosity would have
  been thermalized well below the photosphere, potentially sustaining the
  high luminosity without exhibiting the traditional observational
  signatures of strong CSM interaction (narrow H$\alpha$ emission and
  X-rays).  Variations in density structure of the CSM could account
  for the multiple rebrightenings of the lightcurve.  We propose that a canonical 1$\times$10$^{51}$ erg explosion energy with
  enveloped CSM interaction as seen in some recent SNe, rather than an
  entirely new explosion mechanism, may be adequate to explain the
  peculiar evolution of iPTF14hls.

 \end{abstract}


\section{INTRODUCTION}

The very peculiar Type~II supernova (SN) iPTF14hls was a so-far unique
event.  It had a nearly unchanging and fairly normal spectrum
resembling a traditional Type II-P explosion, but this was difficult
to reconcile with its highly unusual multi-peaked lightcurve, which also
exhibited a high luminosity lasting far longer than any normal
SNe~II-P \citep{2017Natur.551..210A}. The optical lightcurve had a
bumpy plateau lasting over 600 days with 5 separate brightening
episodes, and the velocities in optical spectra evolved 5--10 times
slower than a normal SN~II-P.  Due to the lack of narrow lines of
hydrogen typically seen in interacting Type II SNe (i.e. SNe IIn) and
the absence of any detectable X-ray or radio emission,
\citet{2017Natur.551..210A} argued against shock interaction with
dense circumstellar material (CSM) as the cause for the strangeness of
this SN although they did not definitively rule it out).  Instead it was suggested that some new evolutionary path or
explosion mechanism may need to be invoked to explain the unusual properties
of iPTF14hls \citep{2017Natur.551..210A}.

Based on a number of recent studies of SNe IIn, CSM interaction is
understood to be a pathway to get long lasting, high luminosity, and
irregular or bumpy lightcurves. Through CSM interaction, kinetic
energy of the fast SN ejecta are converted to luminosity in a dense
radiative shock when the SN ejecta slams into the CSM (for a recent
review of interacting SNe, see \citealt{2016arXiv161202006S}).  Dense
shells of CSM can generate extremely high luminosity in some SNe
\citep{2007ApJ...671L..17S}, and some SNe IIn can be powered this
way for a long time \citep{2011ApJ...729...88R} --
much longer than conventional emission from a recombination
photosphere that fades roughly on a diffusion timescale \citep{1996snih.book.....A}.  Since the luminosity generated by CSM
interaction depends on the density of the CSM it overtakes, density
inhomogeneities in the CSM can easily lead to irregular bumpy light
curves.  For instance, \citet{2017A&A...605A...6N} show that iPTF13z
had a similar bumpy lightcurve, but its spectrum was that of a
Type~IIn with strong narrow emission lines.  CSM interaction would
seem a perfectly suitable explanation for the light curve of
iPTF14hls, were it not for the fact that this SN's spectrum showed no
signatures of CSM interaction in spectra taken over the first 600 days
\citep{2017Natur.551..210A}.

If we were restricted to only scenarios that permitted spherical
symmetry, then the lack of narrow lines and X-rays in iPTF14hls would
indeed seem to rule out CSM interaction as a plausible explanation.
However, a diverse range of observations dictate that close binary
interaction \citep{2012Sci...337..444S, 2017ApJS..230...15M}  and
asymmetric CSM \citep{2014ARA&A..52..487S} are the norm among massive stars, not
the exception.  Binary interaction may often lead to asymmetric or
disc-like CSM, especially in cases of eruptive pre-SN mass loss \citep{2014ApJ...785...82S}.  In such cases, the situation becomes more
complicated; over most of the solid angle of the explosion, the fast
SN ejecta expand unimpeded as though there is no CSM.  In
the equator, however, strong CSM interaction occurs as the SN ejecta
are quickly decelerated by the dense CSM.  As such, the slower CSM
interaction region (and hence, the CSM interaction luminosity and most
observable signatures) can be quickly overrun by the rapidly expanding
opaque ionized SN ejecta envelope.  If the CSM interaction region is
buried below the SN ejecta photosphere, then the luminosity created
through the interaction is reprocessed by the opaque SN ejecta. From
most viewing angles off the equator, this might result in a normal
Type~II-P spectrum that lasts through the duration of the main SN
peak.  This sort of evolution has been invoked already in PTF11iqb
\citep{2015MNRAS.449.1876S} as well as SN~1998S, SN~2009ip, and
SN~1993J discussed in \citet[and references
therein]{2016arXiv161202006S}, where this process is referred to as
enveloped or swallowed CSM interaction.  The main condition for
enveloped CSM interaction is that the CSM is highly asymmetric.  While
the narrow and intermediate-width lines can be hidden partly or fully
during the main light curve peak, they tend to emerge again at late
times when the SN fades and the photosphere recedes.  When this
behavior is seen, it suggests that enveloped CSM interaction may have
been contributing to the luminosity the whole time.  In principle, this
mechanism could occur in a wide range of SN subtypes, including SNe
Ia-CSM (see \citealt{2016arXiv161202006S}).

In this paper we show that indeed iPTF14hls has now revealed signs of
strong CSM interaction in optical spectra taken over 3 years after
explosion, and we discuss how this scenario may provide a solution to
the otherwise  puzzling light curve and spectral evolution first
presented by \citet{2017Natur.551..210A}. The paper is structured as
follows: In section 2 we present our new observations, in section 3 we
discuss and interpret these results, and in section 4 we make some
concluding remarks.

\begin{figure}
\includegraphics[width=3.5in]{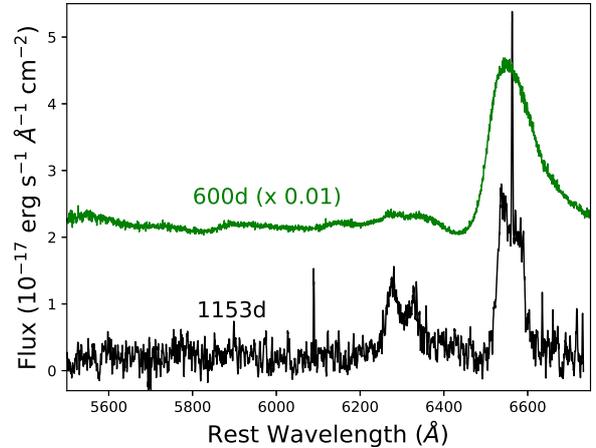}
\caption{Day 1153 (black) and day 600 (green) spectra of
  iPTF14hls. Both have been corrected for redshift and extinction as
  discussed in Section 2.  The day 600 spectrum has also been shifted
  by a constant for clarity and is from \citet{2017Natur.551..210A},
  made available via WISeREP \citep{2012PASP..124..668Y}.}
\label{fig:full}
\end{figure}

\begin{figure*}
\includegraphics[width=5.8in]{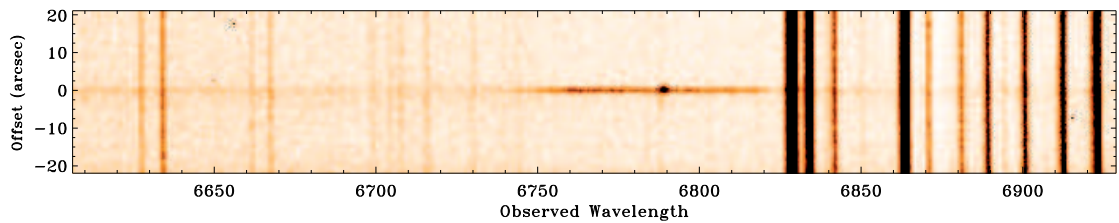}
\caption{Stacked 2D spectrum of iPTF14hls taken on day 1153 after
  discovery, before subtraction of sky or host galaxy light.  The
  intermediate-width H$\alpha$ emission is clearly seen, as is the
  region of the narrow H$\alpha$ emission. There is some very faint
  extended emission to the lower side of the SN that could come from a
  local HII region, but it is at different velocity from that of the
  narrow H$\alpha$ emission.}
\label{fig:2d}
\end{figure*}

\section{OBSERVATIONS}

We obtained 3 $\times$ 1200 s exposures of iPTF14hls with the Blue
Channel spectrograph on the Multiple Mirror Telescope (MMT) on 2017
November 20 UTC (JD 2458077.9). Adopting the date of discovery (JD
2456922) as day 0, following \citet{2017Natur.551..210A}, the date of
our new spectrum corresponds to day 1153. The observations were taken
with the 1200 l mm$^{-1}$ grating with a central wavelength of 6300
\AA\ and a 1$\farcs$0 slit width.  Seeing was 0$\farcs$9.  The
spectral range covers approximately 5700 - 7000 \AA\ as seen in Figure
\ref{fig:full}, and the resolving power (R =
$\lambda$/$\delta\lambda$) is $\sim$ 4500, or $\sim$ 65 km
s$^{-1}$. Standard reductions were carried out using {\tt IRAF}
\footnote{IRAF, the Image Reduction and Analysis Facility is
  distributed by the National Optical Astronomy Observatory, which is
  operated by the Association of Universities for Research in
  Astronomy (AURA) under cooperative agreement with the National
  Science Foundation (NSF)} and wavelength solutions were determined
using internal HeNeAr arc lamps. Flux calibration was achieved using
spectrophotometric standards at a similar airmass taken throughout the
night.

A portion of the combined 2D spectrum which has been bias and
flat-field corrected but not sky subtracted is shown in Figure
\ref{fig:2d}. Both the intermediate-width and narrow H$\alpha$
emission regions are clearly seen.  The narrow H$\alpha$ emission is
spatially unresolved and coincident with the SN; while we cannot rule
out a local H~{\sc ii} region coincident with the SN position, this
narrow emission is not due to spatially extended H~{\sc ii} region
emission in the host. Following \citet{2017Natur.551..210A}, we assume
negligible host-galaxy extinction and correct only for MW extinction
of E(B-V) $=$ 0.0137 \citep{2011ApJ...737..103S}. We adopt a redshift
of $z$ = 0.0333 for the SN, determined from the narrow H$\alpha$
emission (Figure \ref{fig:velocity}, top). This corresponds to a
luminosity distance of 151 Mpc \citep{2016A&A...594A..13P}.

\section{DISCUSSION}

\subsection{Line Evolution}

Figure \ref{fig:full} shows the clear differences in H$\alpha$ and
[O~{\sc i}] line profiles between day 600 and day 1153.  Not only has
the broad H$\alpha$ emission narrowed, but it has taken on a boxy
asymmetrical shape, and a strong narrow emission feature has emerged.
The [O I] $\lambda\lambda$6300,6363 \AA\ line is also much more pronounced
on day 1153 than on day 600.

\begin{figure}
\includegraphics[width=3.3in]{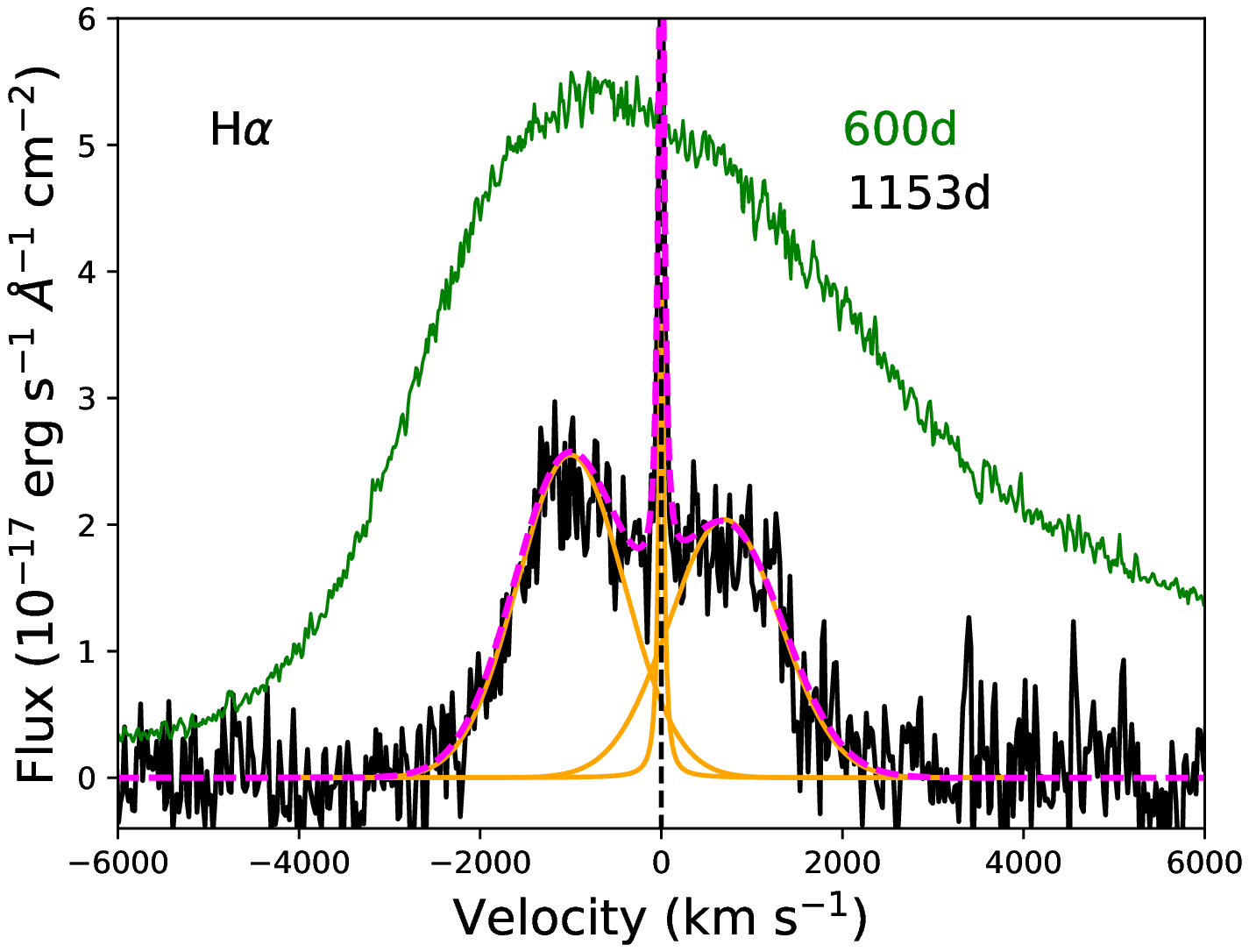}
\includegraphics[width=3.3in]{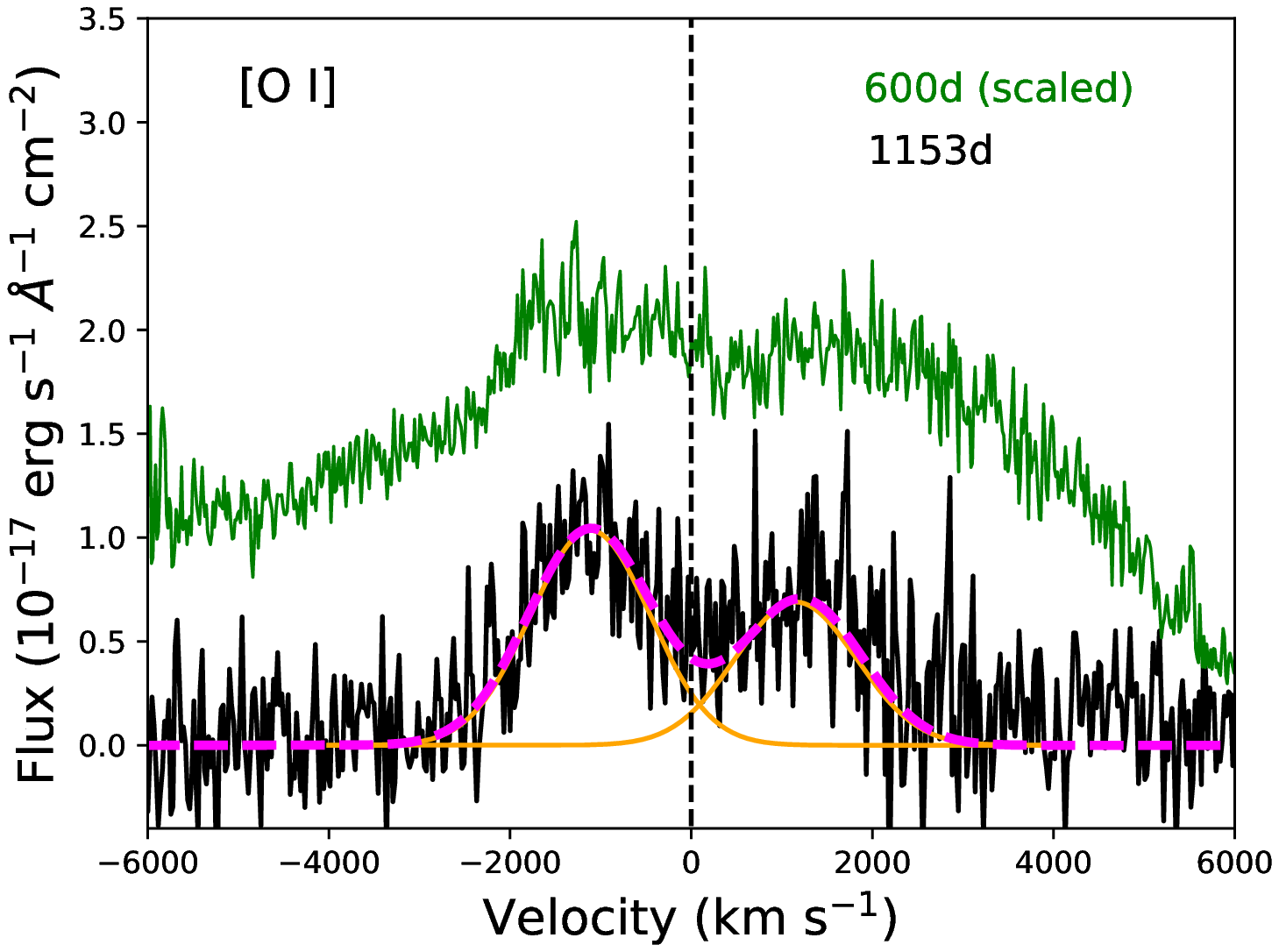}
\caption{Velocity profiles of H$\alpha$ (top) and [O~{\sc i}] (bottom)
  on days 600 and 1153. The individual component fits described in
  Table \ref{tab:fwhm} are shown in orange and the composite in
  magenta, compared to the day 1153 spectrum. }
\label{fig:velocity}
\end{figure}

Inspection of the region around H$\alpha$ in the top panel of Figure
\ref{fig:velocity} shows that the velocity width of H$\alpha$ has
slowed from around 6000 km s$^{-1}$ (day 600) to 3000 km s$^{-1}$ (day
1153).  Structure has also appeared in the once smooth broad line,
with two intermediate-width bumps on the blue and red side of the
narrow emission.  The line profile on day 1153 can be approximated by
a narrow (FWHM = 65 km s$^{-1}$) Lorentzian at zero velocity plus two
Guassians centered at $-$1000 km s$^{-1}$ and 700 km s$^{-1}$, with
FWHM values of 1380 and 1440 km s$^{-1}$, respectively (Table
\ref{tab:fwhm}).  The total H$\alpha$ luminosity is approximately 4
$\times$ 10$^{39}$ erg s$^{-1}$, at least 2 orders of magnitude
fainter than on day 600.   Note that no corresponding photometry were taken with our spectrum to constrain the accuracy of this calibration, so this is a rough approximation.

\begin{table}
\centering
\begin{center}\begin{minipage}{3.3in}
      \caption{FWHM and centres of Emission Components (in km s$^{-1}$) for iPTF14hls on day 1153.}
\begin{tabular}{@{}lccc}\hline\hline
Line&FWHM&centre&Flux \\   
&km s$^{-1}$&km s$^{-1}$&erg s$^{-1}$ cm$^{-2}$\\ \hline
H$\alpha$&1380&-1000&7.8 $\times$ 10$^{-16}$ \\
H$\alpha$&65&0&1.3 $\times$ 10$^{-16}$ \\
H$\alpha$&1440&700&5.8 $\times$ 10$^{-16}$\\
$[$O I$]$&1550&-1110&3.2 $\times$ 10$^{-16}$\\
$[$O I$]$&1570&1150&2.2 $\times$ 10$^{-16}$\\
\hline
\end{tabular}\label{tab:fwhm}
\end{minipage}\end{center}
\end{table}

A similar line-profile shape is seen in the [O~{\sc i}] emission, with
the intermediate-width peaks symmetric about 6300 \AA\ and the profile
becoming slightly flat-topped.  This shape may have been just starting
to emerge in the day 600 spectrum. The peak separation is $\sim$ 46
\AA\, so they are not both blue peaks from the 6300 \AA\ and 6363 \AA\ doublet.  We should also see peaks at 6340 \AA\, and 6386 \AA\ if the 6364 \AA\ behaves similarly.  It is possible they are present, but as the ratio of 6300/6364 is usually 3:1 they would be lost in the noise of the spectrum. Asymmetric oxygen profiles of this type are
discussed in \citet{2010ApJ...709.1343M}, and could be due to dust
either scattering internally or obscuring the red peak.  The stronger
blue peak in H$\alpha$ could also suggest the presence of dust.

Double-peaked profiles in late-time spectra are often interpreted as
due to a disc or torus CSM geometry \citep[for example]{2000ApJ...536..239L,2008ApJ...688.1186H, 2014MNRAS.442.1166M,2016ApJ...832..194K,2015MNRAS.449.1876S, 2017MNRAS.471.4047A}. We strongly suspect that geometry
plays an important role in the profile shape and in the qualitative
behavior of CSM interaction in this SN, as will be discussed in
further detail below.

iPTF14hls is not the first CCSN to show hydrogen line profile shapes
of this type.  Figure \ref{fig:comp} shows the similarities among the
late-time H$\alpha$ emission of interacting SNe~1993J (IIb), 1998S
(IIn), SN2013L (IIn), SN2013ej (II-P/L), 2007od (II-P) and iPTF14hls
\citep{1996ApJ...461..993F,2000AJ....120.1499M,2005ApJ...622..991F,2000ApJ...536..239L,2017MNRAS.471.4047A,2017ApJ...834..118M,
  2010ApJ...715..541A,2011MNRAS.417..261I}.  This shape has also been
seen in other CCSN such as the Type~II-P events SN~2011ja
\citep{2016MNRAS.457.3241A} and SN~2008jb \citep{2012ApJ...745...70P};
the Type~II-L SN~1980K \citep{1990ApJ...351..437F}; and the Type~IIb
SN~2013df \citep{2015ApJ...807...35M}.  While a broad age-range is
represented, the double-peaked or boxy shape and flat-topped profile
can be attributed to interaction with a torus or shell of CSM
\citep{1994ApJ...420..268C}, whether it be from a forward or reverse
shock.  One obvious discrepancy between iPTF14hls and the other SNe is
the width of H$\alpha$ and the velocity of the intermediate-width
peaks.  Both SN 2013ej and SN 1993J have widths extending outwards of
7000$-$9000 km~s$^{-1}$, while iPTF14hls and SN~2013L are much slower
at 1000$-$2000 km s$^{-1}$. Intermediate between the two are SN 1998S
and SN 2007od, with SN 1998S showing no red-peak at all.  This is a
strong indication that the CSM is more massive in the SNe with
narrower profiles, because a higher mass within a given solid angle is
needed to slow a SN ejecta to 1/10th of the initial velocity.  This
relatively high-mass CSM would also be consistent with the high
luminosity observed for iPTF14hls (see below).

\subsection{A possible scenario}

It is a challenge to simultaneously explain the long-lasting (and
bumpy) lightcurve of iPTF14hls while also accounting for the seemingly
unchanging optical spectrum of a normal SN~II-P.  Here we propose that
both may be explained with CSM interaction, providing that a
particular set of conditions are met.  Most importantly, the CSM
must be highly asymmetric, perhaps in a disc or torus, and of limited
radial extent.  As mentioned in the Introduction, and as illustrated
in Figure \ref{fig:sketch}, a disc-like geometry in the CSM may allow
the CSM interaction to be hidden below the photosphere after that disc
is enveloped by the fast SN ejecta.  If the region of CSM interaction
is happening below the ejecta photosphere, it can be hidden for long
periods of time because the sustained CSM interaction luminosity
itself keeps the surrounding SN ejecta ionized and optically thick.
Due to the uncertainty in explosion date of the SN by $\sim$ 100 days,
it is plausible that narrow emission lines were present
at early times before the photosphere swallowed the CSM, as seen in objects
like PTF11iqb. This phenomena is explained extensively in
\citet{2016arXiv161202006S} and \citet{2015MNRAS.449.1876S}, and
requires only that the disc or torus of material has a limited spatial
extent so that it can be overrun early on by the SN photosphere.  The
luminosity and duration of iPTF14hls are more extreme than PTF11iqb,
but the same basic principle may apply.

Since the disc/torus would only intercept a small fraction of the
ejecta, most of the ejecta would expand freely and the ejecta photosphere
would envelope the CSM quickly (Figure \ref{fig:sketch}a). For locations near the equator, the
collision of the ejecta and the dense CSM produces a high luminosity of
X-ray and UV photons that cannot escape due to the opacity of the
surrounding SN ejecta, and are thermalized below the SN ejecta
photosphere.  This causes a temporary increase in the luminosity of
the SN (i.e. a``bump") when a dense shell or clump is encountered by the
shock. The continual internal energy deposition would also keep the spherically expanding ejecta photoionized, causing seemingly static or extremely slow spectral evolution until the photosphere completely recedes.

The reprocessing of CSM interaction luminosity by the
surrounding ejecta not only prevents the X-rays from escaping, but
also smears out any narrow line emission. This would explain why
\citet{2017Natur.551..210A} detected no X-ray emission or narrow lines. Any narrow
emission emitted by pre-shock CSM in the disc would encounter multiple
scatterings throughout the opaque SN ejecta and the extended H$\alpha$
emitting region, effectively blurring the line structure until later
times when the opacity dropped sufficiently. Therefore any narrow line
emitting region could be easily buried within the SN ejecta, and would
not be seen until late times.

\begin{figure}
\includegraphics[width=3.3in]{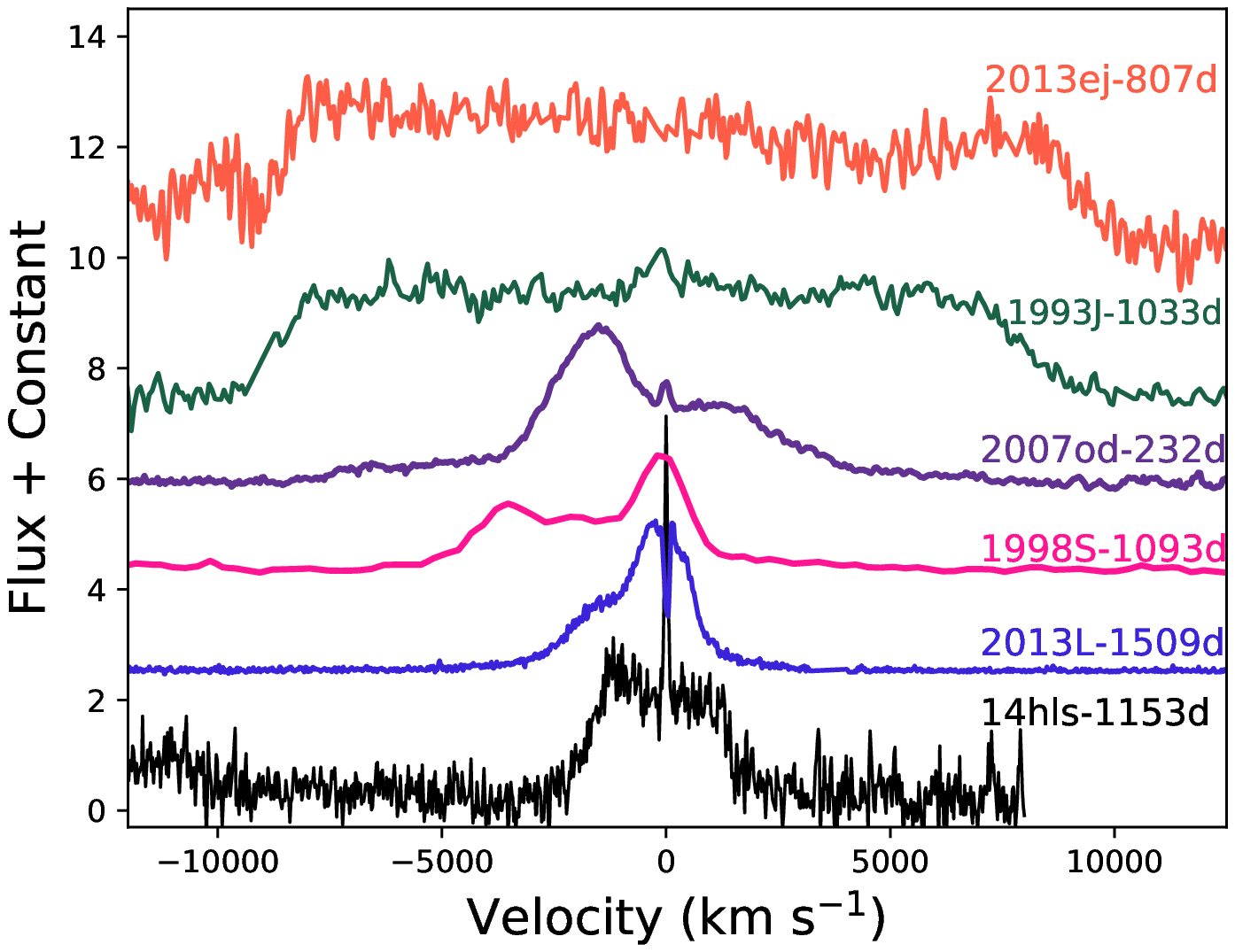}
\caption{The day 1153 H$\alpha$ line of iPTF14hls compared with a few other
  late-time CSM interacting SNe.  Data have been scaled and shifted for
  comparison, and come from \citet[SN2013L]{2017MNRAS.471.4047A},
  \citet[SN1998S]{2004MNRAS.352..457P},
  \citet[SN2007od]{2010ApJ...715..541A},
  \citet[SN1993J]{2014AJ....147...99M}, and
  \citet[SN2013ej]{2017ApJ...834..118M}.}
\label{fig:comp}
\end{figure}

\begin{figure}
\includegraphics[width=3.3in]{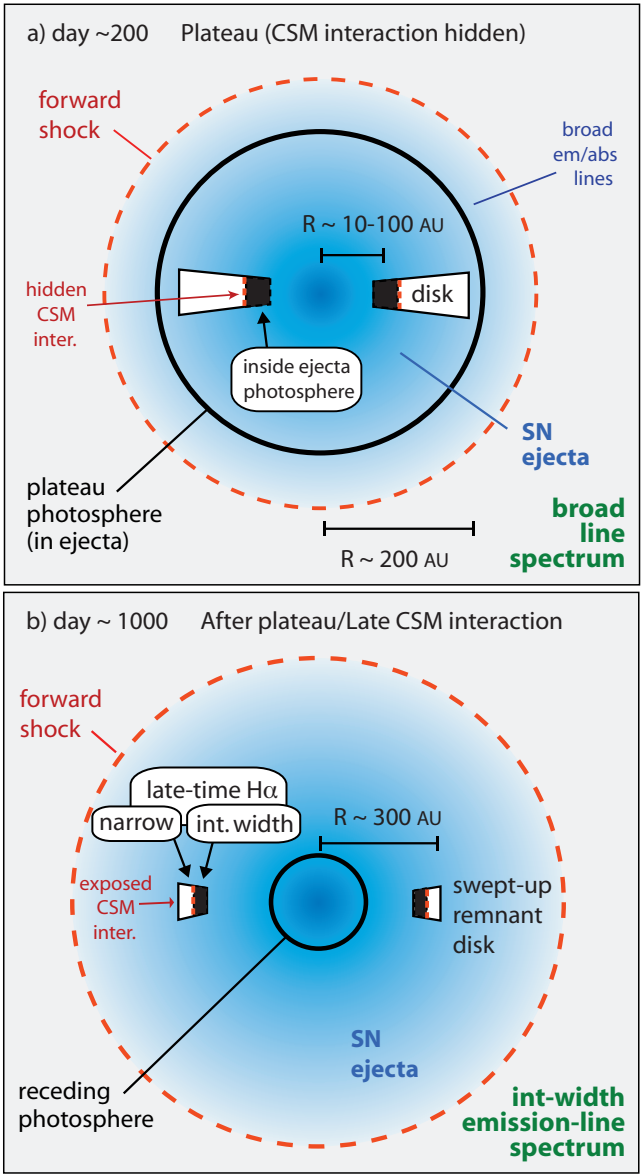}
\caption{A sketch of enveloped CSM interaction, adapted from
  \citet{2015MNRAS.449.1876S}, with a pre-existing disc or torus
  around the progenitor.  (a) At early times the ejecta photosphere
  quickly envelops the CSM.  Strong interaction occurs below the
  photosphere, keeping the luminosity high while hiding observational
  evidence that interaction is occurring.  (b) At late times the
  photosphere has receded and the CSM is revealed again in the form of
  double-peaked intermediate-width spectral lines, as we observe in
  the late-time spectrum of iPTF14hls. Narrow emission arises in the pre-shock disc gas, which is also uncovered at late times.}
\label{fig:sketch}
\end{figure}

Eventually the photosphere will recede (Figure \ref{fig:sketch}b),
revealing the CSM interaction below.  As a consequence of previous shock/CSM
interaction, much of the  $\sim$ 100 km s$^{-1}$ CSM will already have been swept up to speeds
of $\sim$1000 km s$^{-1}$, which would explain the velocity offsets of
the red/blue peaks seen in iPTF14hls.  Any remaining slow moving CSM will be photoionized by the ongoing shock in the disc (red dashed line) and will be observed as narrow H$\alpha$ emission. The remnant of this disc might
also be cool and dense, enough to allow the growth of dust grains that
could attenuate the red peak more than the blue peak.  This has been
seen in numerous Type II SNe, with a few shown in Figure
\ref{fig:comp} \citep{2000ApJ...536..239L, 2009ApJ...695.1334S,
  2009ApJ...691..650F, 2010ApJ...715..541A,2016MNRAS.457.3241A}.

Using the Gaussian fits shown in Table \ref{tab:fwhm}, we estimate the
bulk of the CSM is moving at $\sim$ 1000 km s$^{-1}$.  By the time the
SN had achieved maximum light at day 220, this CSM would have been at
a location of only 1.9 $\times$ 10$^{15}$ cm ( possibly closer if the CSM is not assumed to instantly be accelerated to this velocity.)  This is below the
blackbody photospheric radius of 2 $\times$ 10$^{15}$ cm, fully
consistent with the interpretation of enveloped CSM interaction, and
well below the Fe~{\sc ii} line formation radius of 8 $\times$
10$^{15}$ cm (shown in Figure 4 of \citet{2017Natur.551..210A}).  On
day 600 the intermediate-width peaks would be located at a distance of
$\sim$5$\times$10$^{15}$ cm, outside of the blackbody photosphere, but
still well within the Fe~{\sc ii} line region, which has expanded to
2$\times$10$^{16}$ cm. At all times, the CSM is below the H$\alpha$
line-forming radius, which should be between 5-6 $\times$ 10$^{16}$ cm
by day 1153 assuming an H$\alpha$ velocity of 4000$-$6000 km s$^{-1}$. Since the H$\alpha$ emission from the CSM interaction is still well below the line forming region of the broad H$\alpha$ in the SN ejecta, the internal H$\alpha$ photons would be scattered due to significant line opacity, and the original line shape would be hidden.

 Could the intermediate-width and narrow H$\alpha$ observed at late times arise from the fastest ejecta crashing into a distant CSM shell that had a large inner cavity, so that CSM interaction has just turned on?  That might explain the lack of CSM interaction signatures at early times (although it would provide no help in understanding the peculiar properties of the SN).  This is an unlikely explanation for the origin of the H$\alpha$ emission we report here.  If the fastest ejecta just now encountered a distant hollow shell at a large radius, then in addition to seeing narrow emission from the pre-shock gas and intermediate width emission from the shocked CSM, we would also expect to see very broad emission from the fastest ejecta that are at the reverse shock \citep{1994ApJ...420..268C}.  This is precisely the scenario that occurred in SN1987A, when the blast wave hit distant CSM after a delay of several years, and in that case the broad H$\alpha$ from the reverse shock is quite strong \citep{2005ApJ...635L..41S}.   This broad emission is not seen here, indicating that the shock is not caused by the fastest ejecta intercepting a distant shell, instead, a shock in the CSM at a much smaller radius that was previously hidden below the photosphere would have much slower SN ejecta of only 1000-2000 km s$^{-1}$ catching up to the reverse shock.

The lack of polarization detected by \citet{2017Natur.551..210A}
suggests that the bulk of the continuum emission during the first year or two
of the iPTF14hls light curve is coming from a mostly spherical
environment.  In a simple scenario, this might be at odds with our
requirement of highly asymmetric CSM, as long as the SN is not viewed
pole-on.  However, again, in the scenario we suggest with enveloped
CSM interaction, the asymmetric emitting regions are buried well below
the photosphere.  Since these CSM interaction photons are thermalized
deep inside the SN ejecta envelope, their polarization signature would
be erased.

The asymmetric CSM was likely created through some enhanced episodic
or explosive mass-loss shortly before core-collapse.  If for instance
the CSM is located at a distance between 50-100 AU
(10$^{14.5}$-10$^{15}$ cm; this is the distance derived by
extrapolating the late-time expansion speed of 1000 km s$^{-1}$ to
around day 100$-$200), the fastest narrow H$\alpha$ emission velocity of
$\sim$80 km s$^{-1}$ (measured from the blue edge of the narrow emission line) would require a mass loss episode sometime
between 3 - 6 years before discovery.  This is in agreement with  \citet{2017Natur.551..210A} and the timescale is consistent
with late nuclear burning instabilities such as pulsational pair
instabilities or wave driven mass loss from Ne/O burning
\citep{2012MNRAS.423L..92Q, 2014ApJ...785...82S,
  2017MNRAS.470.1642F,2017ApJ...836..244W}.  In addition to the
advanced burning instabilities, binary interaction with a bloated
envelope would likely need to be invoked in order to explain the
pronounced asymmetry in the CSM \citep{2014ApJ...785...82S}.
 
CSM interaction is an efficient way to convert SN ejecta kinetic
energy into radiative luminosity.  \citet{2017Natur.551..210A}
estimate a total radiated energy for iPTF14hls of about
2$\times$10$^{50}$ erg.  This can easily be achieved with CSM
interaction and a conventional SN explosion energy of 1-2 $\times$
10$^{51}$ erg, provided that a massive disc or torus intercepts about
10-20\% of the solid angle of the SN ejecta \citep{2014MNRAS.438.1191S}.  From momentum conservation, a CSM mass of 5--10
$M_{\odot}$ would be required to decelerate that fraction of a typical
SN ejecta mass down to the final observed coasting speed of 1000 km
s$^{-1}$.  While this CSM mass requires rather extreme pre-SN
mass-loss rates compared to normal steady winds, it is not so extreme
compared to some high-luminosity SNe~IIn such as SN~2006gy, SN~2006tf,
and SN2010jl (see \citet{2014ARA&A..52..487S} for a review of pre-SN mass loss).  This
sort of scenario is also less extreme than the 50 $M_{\odot}$ or more of SN
ejecta, unusually high explosion energy of 10$^{52}$ erg, and perhaps multiple explosions, required in a scenario without CSM interaction
\citep{2017Natur.551..210A}.

\section{Conclusions}

While the evolution of iPTF14hls is quite unusual among SNe studied so
far,  we propose that it may be explained with an enveloped CSM interaction scenario.  With two assumptions --- (1) that
the SN progenitor had an extreme mass-loss episode a few years before
explosion, as is generally thought to be the case for most SNe IIn,
and (2) that binary interaction or rapid rotation caused this CSM to
be distributed into a disc or torus --- it may not be so challenging
to explain both the observed spectrum and light curve of iPTF14hls, as
well as the multi-peaked and boxy emission-line profiles seen at late
times.  This is possible if the CSM is dense and close enough to the
star that it can be overrun and enveloped by the SN ejecta.  The CSM
interaction region could be hidden within the SN photosphere for
months or years, but its luminosity would be thermalized and would increase the overall
optical luminosity of the event by keeping the surrounding SN ejecta
ionized.  Viewed from an outside observer, the SN ejecta that are
reheated by CSM interaction from deep inside would exhibit the
spectrum of a normal, broad Type II-P event. Multiple shells or some
other density inhomogeneities in the CSM could cause the emergent
light curve to appear bumpy as the shock propagates through each of
them. Therefore it is plausible that the strange behavior of iPTF14hls
is due to circumstellar interaction, despite the lack of obvious
signatures of CSM interaction in the early spectra.

 There may indeed have been a brief period of time when iPTF14hls showed strong narrow lines and other signs of CSM interaction, but because the first spectra were obtained around 100 days after explosion, these might have already faded when the CSM interaction region was enveloped by the opaque fast SN ejecta. Additional observations can help constrain the mass-loss history of
the progenitor and further uncover the interaction occurring in
iPTF14hls.  For example, although early time X-ray and radio
observations yielded non-detections, the current interaction seen in
H$\alpha$ may suggest that deep late-time X-ray observations
should be obtained.  The IIn SN 2005ip had an X-ray luminosity  $\sim$25$\times$ brighter than the H$\alpha$ luminosity at late times \citep{2017MNRAS.466.3021S}. Assuming iPTF14hls  behaves similarly we can estimate an L$_{X}$ = 1$\times$10$^{41}$ erg s$^{-1}$. At the distance of iPTF14hls and similar assumptions about absorption as in SN 2005ip,  Chandra ACIS observations with roughly an hour of on source time should readily detect this. It would also be interesting to produce 2-D radiative
transfer models for this asymmetric scenario of enveloped CSM
interaction, to see exactly what CSM geometries and viewing angles are consistent
with a lack of narrow lines in iPTF14hls.

\smallskip\smallskip\smallskip\smallskip
\noindent {\bf ACKNOWLEDGMENTS}
\smallskip \footnotesize Special thanks to Jon Mauerhan for the
spectrum of SN 2013ej and to the referee for constructive comments.  N.S. acknowledges puzzling but fruitful
conversations with I. Arcavi about iPTF14hls. The day 600 spectrum of
iPTF14hls was obtained from the Weizmann interactive supernova data
repository (http://wiserep.weizmann.ac.il). Support was provided by
NSF grants AST-131221 and AST-151559, and by a Scialog grant from the
Research Corporation for Science Advancement.

\bibliographystyle{mnras}
\bibliography{iPTF14hls}
\bsp	
\label{lastpage}
\end{document}